\documentclass[12pt]{article}
\usepackage{graphicx}
\usepackage{latexsym}
\usepackage{amssymb}
\usepackage{amsmath}
\usepackage{amsthm}
\usepackage{stmaryrd}
\usepackage[latin1]{inputenc}
\usepackage[english]{babel}
\usepackage{feynmf}
\usepackage{fancyhdr}
\usepackage{indentfirst}
\usepackage{float}
\newcommand{\sh}[1]{#1\hskip-7pt \diagup}
\begin{document}

\begin{center}
\Large{\bf Radiative Corrections in a Vector-Tensor Model}

\bigskip

\large{\bf F. Chishtie, M. Gagn\'e-Portelance, T. Hanif,\\
S. Homayouni, D. G. C. McKeon}\\

Department of Applied Mathematics, The University of Western Ontario\\
London, Ontario N6A 5B7  CANADA
\end{center}

\bigskip

\noindent Fax: (519) 661-3523\\
\noindent Tel:  (519) 661-2111, ext. 88789\\
\noindent E-mail:  dgmckeo2@uwo.ca \hfill PACS No: 11.10.q\\

\begin{abstract}
In a recently proposed model in which a vector non-Abelian gauge field 
interacts with an antisymmetric tensor field, it has been shown that 
the tensor field possesses no physical degrees of freedom.  This formal 
demonstration is tested by computing the one-loop contributions of the 
tensor field to the self-energy of the vector field.  It is shown that 
despite the large number of Feynman diagrams in which the tensor field 
contributes, the sum of these diagrams vanishes, confirming that it is 
not physical.  Furthermore, if the tensor field were to couple with a 
spinor field, it is shown at one-loop order that the spinor self-energy 
is not renormalizable, and hence this coupling must be excluded.  In 
principle though, this tensor field does couple to the gravitational 
field.
\end{abstract}

Recently, a model has been considered in which a non-Abelian gauge 
field $W_\mu^a$ interacts with an antisymmetric tensor field 
$\phi_{\mu\nu}^a$ with the Lagrange density \cite{1}
\begin{eqnarray}%1
L = -\frac{1}{4} F_{\mu\nu}^a F_{\mu\nu}^a & + & \frac{1}{12} 
G_{\mu\nu\lambda}^a G_{\mu\nu\lambda}^a  + \frac{m}{4} 
\epsilon_{\mu\nu\lambda\sigma} \phi_{\mu\nu}^a F_{\lambda\sigma}^a\nonumber\\
& + & \frac{\mu^2}{8} \epsilon_{\mu\nu\lambda\sigma} \phi_{\mu\nu}^a 
\phi_{\lambda\sigma}^a \; .
\end{eqnarray}

In eq. (1), $m$ and $\mu$ are mass parameters,
\begin{equation}%2
F_{\mu\nu}^a = \partial_\mu W_\nu^a - \partial_\nu W_\mu^a + gf^{abc} 
W_\mu^b W_\nu^c
\end{equation}
\begin{equation}%3
G_{\mu\nu\lambda}^a = D_\mu^{ab} \phi_{\nu\lambda}^b + D_\nu^{ab} 
\phi_{\lambda\mu}^b + D_\lambda^{ab} \phi_{\mu\nu}^b
\end{equation}
and
\begin{equation}%4
D_\mu^{ab} = \partial_\mu \delta^{ab} + gf^{apb} W_\mu^p \; .
\end{equation}

This lagrange density is invariant under the infinitesmal gauge 
transformation
\begin{equation}%5
\delta W_\mu^a = D_\mu^{ab}\Omega^b \; \; \; \; \delta 
\phi_{\mu\nu}^a = gf^{abc}\phi_{\mu\nu}^b \Omega^c \; .
\end{equation}
Both by a canonical analysis using the Dirac constraint formalism 
\cite{2} and by explicit eliminations of non-physical degrees of freedom, it 
has been shown that in the Abelian limit, the tensor field in eq. (1) 
does not possess any physical degrees of freedom.

It was surmised that the full non-Abelian model of eq. (1) also does 
not contain any physical degrees of freedom associated with the tensor 
field;  this conjecture is what we can test by an explicit calculation.  
Evaluation of the one-loop contributions to the vector self-energy $\langle W^{a}_{\mu} W^{b}_{\nu} \rangle$
 involves Feynman diagrams with both vertices and 
propagators associated with the tensor field, and if the tensor field is 
indeed non-physical, its contributions to this Green's function should 
all cancel.  Below we show that this is in fact what happens.

Working in Euclidean space, the contribution to $L$ in eq. (1) that is 
bilinear in the fields is

\begin{equation}%6
L^{(2)} = \frac{1}{2} (W_\lambda, \phi_{\alpha\beta})
\left( 
\begin{array}{clcr}\partial^2 I_{\lambda\sigma} & \frac{m}{2} 
B_{\lambda, \gamma\delta} \\
\frac{m}{2} A_{\alpha\beta, \sigma} & -\frac{1}{2}I_{\alpha\beta, 
\gamma\delta} \partial^2 + Q_{\alpha\beta, \gamma\delta} + \frac{\mu^2}{4} 
\epsilon_{\alpha\beta\gamma\delta}
\end{array}
\right)
\left( 
\begin{array}{clcr} W_\sigma \\ \phi_{\gamma\delta}
\end{array} 
\right)
\end{equation}
where
\begin{eqnarray}%7
I_{\alpha\beta} & = & \delta_{\alpha\beta}, \; \; L_{\alpha\beta} = 
\partial_\alpha \partial_\beta, \; \; I_{\alpha\beta,\gamma\delta} = 
\frac{1}{2}\left( \delta_{\alpha\gamma} \delta_{\beta\delta} - 
\delta_{\alpha\delta} \delta_{\beta\gamma}\right)\nonumber\\
A_{\mu\nu , \lambda} & = & \epsilon_{\mu\nu\kappa\lambda} 
\partial_\kappa = -B_{\lambda, \mu\nu}\nonumber\\
Q_{\mu\nu, \lambda\sigma} & = & \frac{1}{4} \left( \delta_{\mu\lambda} 
\partial_{\nu\sigma}^2 + \delta_{\nu\sigma} \partial_{\mu\lambda}^2 - 
\delta_{\mu\sigma} \partial_{\nu\lambda}^2
- \delta_{\nu\lambda} \partial_{\mu\sigma}^2\right)\nonumber \\
L_{\mu\nu, \lambda\sigma} & = & \epsilon_{\mu\nu, \kappa\lambda} 
\partial_\kappa \partial_\sigma - \epsilon_{\mu\nu \kappa \sigma} 
\partial_\kappa \partial_\lambda = - R_{\lambda\sigma, \mu\nu}\; .
\end{eqnarray}

We have used a gauge fixing Lagrangian
\begin{equation}%8
L_{gf} = -\frac{1}{2} (\partial \cdot W^a)^2 .
\end{equation}

The inverse of the matrix $M$ appearing in eq. (6) is
\begin{equation}%9
M^{-1} = 
\left(
\begin{array}{clcr}
\frac{I_{\sigma\kappa}}{\partial^2} & -\left(\frac{m}{\mu^2 
\partial^2}\right) D_{\sigma, \pi\tau}\\ \\
\left( \frac{m}{\mu^2 \partial^2}\right) C_{\gamma\delta, \kappa} & 
\left( \frac{4}{\mu^4}\right) \left( 1 - \frac{m^2}{\partial^2}\right) 
Q_{\gamma\delta, \pi\tau} + \left( \frac{1}{\mu^2 \partial^2}\right) 
\left( -L_{\gamma\delta, \pi\tau} + R_{\gamma\delta, \pi\tau}
\right)
\end{array}
\right)  
\end{equation}
(This corrects a minor mistake in ref. \cite{1}.)

Using eq. (9), the free field propagators can be determined.  The 
Feynman rules needed to determine the contribution of the tensor field to $\langle W^{a}_{\mu} W^{b}_{\nu} \rangle$
 are in fig. 1.  

In computing the one-loop corrections to $\langle W^{a}_{\mu} W^{b}_{\nu} \rangle$, it is 
necessary to include diagrams in which the external leg involves the 
mixed propagator $\langle W_\mu^a \phi_{\lambda\sigma}^b \rangle$.

The presence of the tensor $\epsilon_{\mu\nu\lambda\sigma}$ in the 
Lagrange density of eq. (1) makes straightforward application of 
dimensional regularization difficult.  The aspects of dimensional regularization 
needed are that shifts of variables of integration in Feynman integrals 
do not generate surface terms, that massless tadpole integrals of the 
form $\int \frac{d^{n} k}{(2\pi)^n} \frac{1}{(k^2)^a}$ vanish, and that
\begin{equation}%10
\epsilon_{\alpha\beta\gamma\delta} \epsilon_{\mu\nu\lambda\sigma} = 
(\delta_{\alpha\mu} \delta_{\beta \nu} \delta_{\gamma\lambda} 
\delta_{\nu\sigma} + \ldots )
\end{equation}

\noindent where in eq. (10) all 24 terms formed by permutting indices 
(taking into account the antisymmetry of 
$\epsilon_{\alpha\beta\gamma\delta}$) are taken into account.  We then use the $n$ dimensional 
relations $\delta_{\mu\mu} = n$ and
\begin{equation}%11
\int \frac{d^n k}{(2\pi)^n} k_\mu k_\nu f(k^2) = \frac{1}{n} 
\delta_{\mu\nu} \int \frac{d^n k}{(2\pi)^n} k^2 f(k^2).
\end{equation}
It turns out though that no integral over loop momentum has to be performed.

The Feynman diagrams associated with the one-loop corrections to $\langle W^{a}_{\mu} W^{b}_{\nu} \rangle$
 involving the tensor field all vanish individually 
except for the one of fig. (2).  These are individually non-zero, but their 
sum reduces to an integral of the form
\begin{equation}%12
\Pi_{\mu\nu} (p) = \int \frac{d^n k}{(2\pi)^n} \int_0^1 dx (1-2x) 
f(x(1-x), k^2, p^2) (p^2 \delta_{\mu\nu} - p_\mu p_\nu).
\end{equation}
On account of the integral over $x$, this too vanishes;  we thus 
conclude that only the usual Yang-Mills diagrams contribute to $\langle W^{a}_{\mu} W^{b}_{\nu} \rangle$
 at one-loop order.  This result is consistent with the 
conclusion reached in ref. \cite{1} that there are no dynamical degrees of 
freedom associated  with $\phi_{\mu\nu}^a$.

In view of this peculiar feature of the tensor field, it is interesting 
to examine how it might couple to matter.  Let us consider a spinor
field $\psi$ and suppose that there are two interaction forms in 
addition to those of eq. (1),
\begin{equation}%13
L_1 = g \bar{\psi} \gamma_\mu \tau^a W_\mu^a \psi
\end{equation}
a\begin{equation}%14
L_2 = h \bar{\psi} \sigma_{\mu\nu} \tau^a \phi_{\mu\nu}^a \psi
\end{equation}
where $\sigma_{\mu\nu} = \frac{1}{2} [\gamma_\mu , \gamma_\nu]$.  One 
can now examine the one-loop corrections to the self-energy of the 
spinor $\langle \psi \bar{\psi}\rangle$.  It turns out the radiative correction 
proportional to $h^2$ and $gh$ contains divergences proportional to
\begin{equation}%15
D_1 = h^2 \left( \frac{m^2}{\mu^4} \right) p^2 \sh{p}
\end{equation}
and
\begin{equation}%15
D_2 = gh \left( \frac{m}{\mu^2} \right) p^2 
\end{equation}
respectively.  Neither of these divergences are consistent with 
renormalizability, and so the tensor-spinor interaction of eq. (14) must be 
excluded, much as we cannot incorporate the magnetic moment interaction 
$\bar{\psi} \sigma_{\mu\nu} \tau^a F_{\mu\nu}^a \psi$ into the Lagrange 
density.

One might well ask what the role of the tensor field might be, seeing 
as its apparent coupling in eq. (1) to the vector field appears to have 
no physical effect.  However, working from the principle that nothing 
that is forbidden is allowed, we cannot exclude its presence.  If 
$\phi_{\mu\nu}^a$ were to exist though, it would necessarily couple to the 
gravitational field, contributing to so-called ``dark matter".

\bigskip

Discussions with S. V. Kuzmin and N. Kiriushcheva are gratefully 
acknowledged.  NSERC provided financial support.  R. and D. MacKenzie were 
helpful.

\eject

%\vspace*{19.0cm}
[1]
\begin{center}
\includegraphics{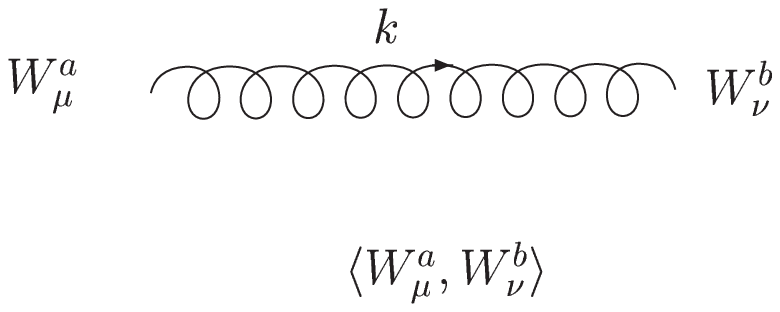}
\end{center}
\begin{eqnarray*}
\frac{\delta_{\mu\nu}\delta^{ab}}{k^2}
\end{eqnarray*}

[2]
\begin{center}
\includegraphics{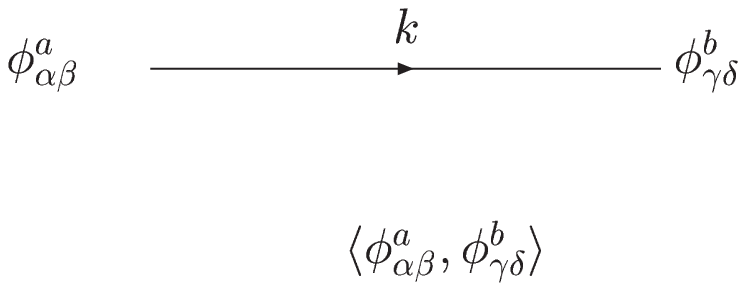}
\end{center}
\begin{eqnarray*}
\frac{\delta^{ab}}{\mu^4} \left( 1 + \frac{m^2}{k^2} \right) \left( 
\delta_{\alpha\gamma} k_\beta k_\delta - \delta_{\beta\gamma} k_\alpha 
k_\delta + \delta_{\beta\delta} k_\alpha k_\gamma - \delta_{\alpha\delta} 
k_\beta k_\gamma\right)\\
+ \frac{\delta^{ab}}{\mu^2 k^2} \left( 
\epsilon_{\alpha\beta\lambda\gamma} k_\delta - \epsilon_{\alpha\beta\lambda\delta} k_\gamma + 
\epsilon_{\gamma\delta\lambda\alpha} k_\beta - 
\epsilon_{\gamma\delta\lambda\beta} k_\alpha \right) k_\lambda
\end{eqnarray*}

[3]
\begin{center}
\includegraphics{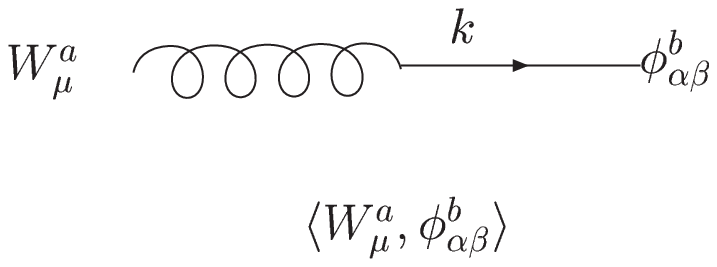}
\end{center}
\begin{eqnarray*}
-\frac{i m \delta^{ab}}{\mu^2 k^2} \left( \delta_{\alpha\mu} k_\beta - 
\delta_{\beta\mu} k_\alpha\right)
\end{eqnarray*}

[4]
\begin{center}
\includegraphics{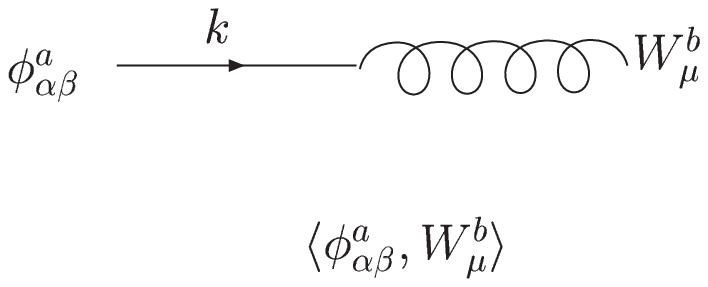}
\end{center}
\begin{eqnarray*}
\frac{i m \delta^{ab}}{\mu^2 k^2} \left( \delta_{\alpha\mu} k_\beta - 
\delta_{\beta\mu} k_\alpha \right)
\end{eqnarray*}

[5]
\begin{center}
\includegraphics{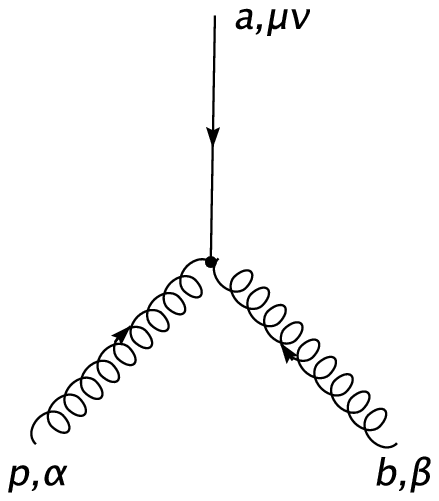}
\end{center}
\begin{eqnarray*}
\frac{m}{2} g f^{abc} \epsilon_{\mu\nu\alpha\beta}
\end{eqnarray*}

[6]
\begin{center}
\includegraphics{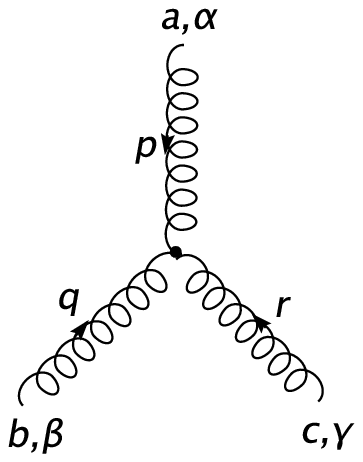}
\end{center}
\begin{eqnarray*}
i g f^{abc} \left[ (p-q)_\gamma \delta_{\alpha\beta} + (r - p)_\beta 
\delta_{\gamma\alpha} + (q-r)_\alpha \delta_{\beta\gamma} \right]
\end{eqnarray*}

[7]
\begin{center}
\includegraphics{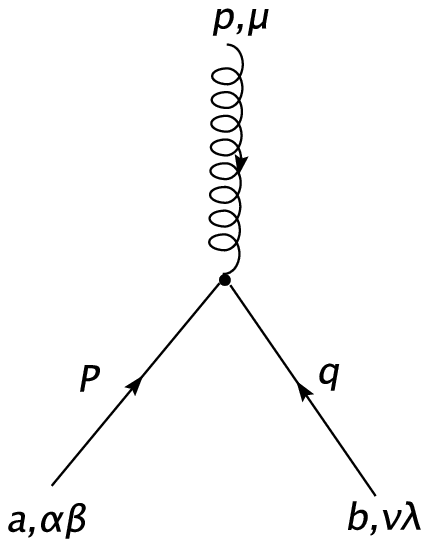}
\end{center}
\begin{eqnarray*}
\frac{i}{4} g f^{apb} \left[ (p-q)_\mu (\delta_{\alpha\nu} 
\delta_{\beta\lambda} - \delta_{\alpha\lambda} \delta_{\beta\nu}) + p_\nu 
(\delta_{\alpha\lambda} \delta_{\mu\beta} - \delta_{\beta\lambda} 
\delta_{\mu\alpha})\right. \\
\left. - p_\lambda (\delta_{\alpha\nu} \delta_{\mu\beta} - 
\delta_{\beta\nu} \delta_{\mu\alpha}) - q_\alpha
(\delta_{\mu\lambda} \delta_{\beta\nu} - \delta_{\mu\nu} 
\beta_{\beta\lambda}) + q_\beta (\delta_{\mu\lambda} \delta_{\alpha\nu} - 
\delta_{\mu\nu} \delta_{\alpha\lambda} \right] 
\end{eqnarray*}
\begin{center}
\bf Feynman Rules\\
 Fig. 1
\end{center}

\begin{figure}[H]
\begin{center}
\includegraphics{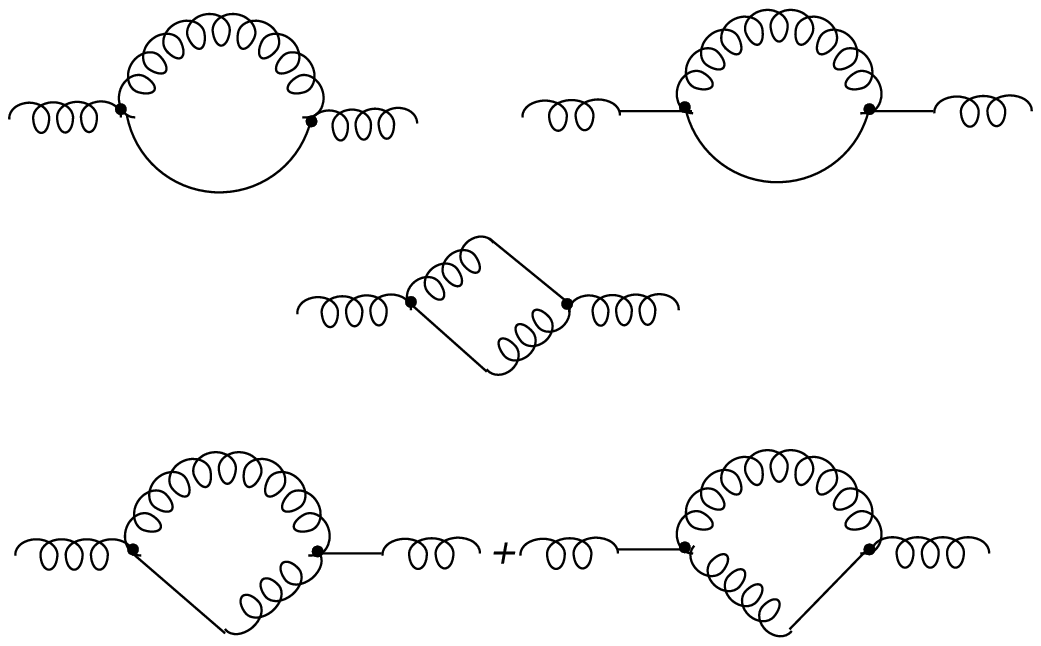}
\end{center}
\end{figure}
\begin{center}
\bf Non-Vanishing Contributions to $\langle W^{a}_{\mu} W^{b}_{\nu} \rangle$ \\
Fig. 2
\end{center}

\end{document}